# Time-frequency analysis of Transitory/Permanent frequency decrease in civil engineering structures during earthquakes

Clotaire MICHEL[1], Philippe GUEGUEN[1,2]

[1] LGIT, University of Grenoble, France

[2] LCPC, Paris, France

Corresponding author

C. Michel

LGIT

BP 53

38041 Grenoble cedex 9

France

cmichel@obs.ujf-grenoble.fr

Tel: +33 4 76 82 80 71

Fax: +33 4 76 82 81 01

Abstract

The analysis of strong motion recordings in structures is a key point in order to understand the damaging process during earthquakes. One of the interesting representations of these signals is in the time-frequency plane. Using the reassigned smoothed pseudo-Wigner-Ville method, a very precise method, we followed the variation in time of the resonance frequencies of the R. Millikan Library (Pasadena, California) and the Grenoble City Hall building (France) during earthquakes. Under strong motions like the San Fernando Earthquake, a quick frequency drop followed by a slower increase is shown and attributed to a loss of stiffness, i.e. damage, of the soil-structure system. However, in the case of weak earthquakes recorded in the Grenoble City Hall building, we show that the greatest variations are due to variation of the ground motion frequency content. Therefore, they cannot be interpreted in terms of variation of the parameters of the soil-structure system in that case.

Introduction

Since Omori [1922], certainly the first author that related the variations of the fundamental frequency in buildings with damage and structural retrofitting, an abundant and recent scientific literature (e.g. [Mucciarelli et al., 2004], [Clinton et al., 2006], [Dunand et al., 2006], [Zembaty et al., 2006]) has shown how these variations can be monitored. The damaging process in buildings during earthquakes produces a permanent loss of structural stiffness and then a permanent decrease of the fundamental frequency, introducing the non-linear modelling of buildings by the earthquake engineering community. Moreover, Clinton et al. [2006] and Dunand et al. [2006] showed also transient decreases of frequency due to the opening and closing process of pre-existing cracks in the structure. Other authors ([Clinton et al., 2006], [Todorovska and Al Rjoub, 2006]) recently showed that small frequency variations under weak and strong motion could also be due to the elastic property variations of the soil-structure system. For earthquake engineering purposes, the physical meaning of the instantaneous variation is a crucial point that must be explored in order to relate it to the properties of the buildings to fix the structural models.

This variation can be studied using the time-frequency representation [Neild et al., 2003] applied to earthquake recordings in buildings, in order to follow the frequency energy distribution in time and to relate it to the amplitude of the shaking [Bradford, 2006]. In this paper, we used the smoothed reassigned pseudo-Wigner-Ville method, a very precise method both in time and frequency. It has been applied to strong motions recorded in the R. Millikan Library (Pasadena, California) and weak earthquakes recorded in the Grenoble City Hall building (France). The origin of the variations observed in buildings is finally discussed.

1. Analysis Method: reassigned Wigner-Ville Time Frequency Distribution

Numerous time-frequency methods exist in the literature, with application to engineering structures (e.g., [Trifunac et al., 2001], [Hans et al., 2000], [Argoul and Le, 2004], [Bradford, 2006], [Todorovska and Trifunac, 2007]). Neild et al. [2003] reviewed these methods for structural dynamics application. They divided them into three categories: the instantaneous frequency, the windowed methods and the bilinear distributions, working simultaneously in time and frequency.

Let define $x_a(t)$ the analytical signal of $x(t)$ by:

$$x_a(t) = x(t) + j\hat{x}(t) \quad (1)$$

where $\hat{x}(t)$ the Hilbert Transform of $x(t)$ is defined by:

$$\hat{x}(t) = \frac{1}{\pi} \int_{-\infty}^{\infty} \frac{x(\tau)}{t - \tau} d\tau \quad (2)$$

$x_a(t)$ can be rewritten in its polar form:

$$x_a(t) = E(t)e^{j\varphi(t)} \qquad (3)$$

with E(t) and φ(t) the envelope and the instantaneous phase of the signal, respectively. For all the time-frequency methods, the analytical signal $x_a(t)$ is generally used instead of the real signal x(t) [Neild et al., 2003].

The windowed methods are the most widely used. The spectrogram consists in computing Fourier Transforms on short time windows. The use of the spectrogram needs a compromise between the frequency and the time precision. The length of the window is also a crucial point with strong influence on the smoothing of the results. In order to avoid this trade-off between the frequency and the time precision, time-frequency distributions are used. They were first developed by signal processors and they are called "energy distributions" because they distribute the energy of the signal in the time-frequency space [Auger et al., 1995]. Most of them used the Cohen's class distribution ([Auger et al., 1995], [Neild et al., 2003]), which keeps the total energy, the instantaneous energy and the spectral density energy of the signal.

The simplest distribution of Cohen's class is called the Wigner-Ville distribution ($P_{WV}$):

$$P_{WV}(t,f) = \int_{-\infty}^{\infty} e^{-j2\pi f\tau} x_a\left(t+\frac{\tau}{2}\right) x_a^*\left(t-\frac{\tau}{2}\right) d\tau \qquad (4)$$

The integral in time is evaluated from -∞ to +∞ and not only on a short time window like the spectrogram so that there is no limit on the frequency precision.

In practice, the smoothed pseudo-Wigner-Ville distribution $P_{spWV}$ is used, corresponding to the windowed version in time and frequency of the $P_{WV}$. It is equivalent to a time and frequency smoothing [Auger et al., 1995] and is adapted to finite signals in time. This distribution is expressed as follows:

$$P_{spWV}(t,f;g,h) = \int_{-\infty}^{\infty} h(\tau) \int_{-\infty}^{\infty} g(u-\tau) e^{-j2\pi f\tau} x_a\left(u+\frac{\tau}{2}\right) x_a^*\left(u-\frac{\tau}{2}\right) du\, d\tau \qquad (5)$$

with *g* and *h* regular windows (for example Hanning windows) corresponding to time and frequency smoothing, respectively.

The drawback of the $P_{spWV}$, is that terms are added in the distribution when we compute the product $x_a\left(t+\frac{\tau}{2}\right)x_a^*\left(t-\frac{\tau}{2}\right)$ producing interferences. In addition, the energy band of the distribution is generally thick, so that it is difficult to interpret the frequency variation. In order to remove these difficulties, the reassignment method can be used [Auger et al., 1995]. These authors assume there are no physical reasons for the energy distribution to be symmetric at the vicinity of each time-frequency point. The principle of the reassignment method is to reassign the energy of the local energy distribution to the centre of gravity of the domain around each time-frequency point. The reassigned distribution is the sum of the distribution obtained in each point. In a mathematical viewpoint, the reassigned smoothed pseudo-Wigner-Ville distribution ($P_{rspWV}$) can be written as follows [Auger et al., 1995]:

$$P_{rspWV}(t',f';g,h) = \int_{-\infty}^{\infty} P_{spWV}(t,f;g,h)\delta(t'-\hat{t}(t,f))\delta(f'-\hat{f}(t,f))dtdf \qquad (6)$$

with

$$\begin{cases} \hat{t}(t,f) = t - \dfrac{P_{spWV}(t,f;t \times g,h)}{2\pi P_{spWV}(t,f;g,h)} \\ \hat{f}(t,f) = f + j\dfrac{P_{spWV}(t,f;t \times g,\frac{dh}{dt})}{2\pi P_{spWV}(t,f;g,h)} \end{cases}$$

In the case of earthquake recordings in buildings, the reassigned smoothed pseudo-Wigner-Ville distribution is precise enough in time and frequency to get the low frequency variations generally assumed for the buildings. In order to avoid the effect of large amplitudes as observed in buildings during earthquakes, the signal is made stationary by dividing it by its envelope, previously defined using the analytic signal (Eq. 3).

The Time-Frequency ToolBox or TFTB [Auger et al., 1995], a compilation of Matlab and Octave scripts under GNU license, was used for this study.

2. Strong motions – Application to the Millikan Library

Clinton et al. [2006], Bradford [2006] and Dunand et al. [2006] showed the permanent and transient drops of the resonance frequencies of the Millikan Library on the Caltech campus (Pasadena California) since its construction. It is a 9-story reinforced-concrete (RC) structure built in 1967 with 21 m wide, 22.9 m long and 43.9 m high. The structural system is based on RC frames and two RC-shear walls in the N direction. A core of RC shear wall adds stiffness to the structure. A harmonic shaker is permanently installed at the top of the structure since the beginning of the seventies to perform forced vibration tests [Kuroiwa, 1967]. Thanks to the Californian Strong motion Instrumentation Program (CSMIP), the building has been instrumented since 1968 with two 3C accelerometers at the top and the basement. After the construction of the building, forced vibration tests were performed in 1967 and the first resonance frequencies were 1.45 Hz and between 1.91 and 1.98 Hz in the E-W and N-S directions, respectively ([Kuroiwa, 1967], [Clinton et al., 2006]).

On February the 9$^{th}$ 1971, the San Fernando earthquake ($M_L$=6.6 at 31 km) produced peak top accelerations of 306 and 341 cm/s$^2$ in the E-W and N-S directions, respectively, one of the greatest recorded in the structure [Clinton et al., 2006], with a peak ground acceleration of 2 m/s$^2$. This earthquake induced cracking and spalling of the concrete slabs on the ground floor and horizontal cracks in the core shear walls between the basement and the second story in the N-S direction ([Foutch and Jennings, 1978], [Clinton et al., 2006]).

The reassigned smoothed pseudo-Wigner-Ville distribution has been calculated for recordings of this earthquake at the top of the building (Fig. 1). In the E-W direction, Figure 1 shows a fast decrease of the first frequency during 15 s between a pre-seismic frequency and a minimum value called here co-seismic frequency. The pre-seismic frequency cannot be seen on this figure because of the too short pre-event time window but it is greater than 1.3 Hz. The co-seismic frequency is 0.94 Hz, i.e. a transitory drop of 35% with respect to the pre-seismic frequency value (1.45 Hz) given by Clinton et al. [2006]. They give a co-seismic frequency of 1 Hz, a bit larger than the value found here. This co-seismic frequency occurs 5 s after the peak acceleration. Once this value is reached, a slow increase starts up to the frequency at the end of the recording, called here post-seismic frequency and equal to 1.15 Hz. This post-seismic frequency is close to the value obtained by forced and weak vibration tests in 1974 (1.21 Hz) [Clinton et al., 2006]. That means that the greatest part of the transient stiffness drop is recovered in 50 s.

In N-S direction, the frequency variation is the same as in the E-W direction, i.e. a fast frequency decrease (10 s) followed by a slow increase (Fig. 1). The pre-seismic frequency is not well defined as well but greater than 1.7 Hz. The co-seismic frequency is 1.52 Hz, i.e. a 20% drop from the pre-seismic value (1.9 Hz) [Clinton et al., 2006]. These authors give a co-seismic frequency of 1.64 Hz, once again a bit greater than the value displayed Fig. 1. The post-seismic frequency value found here (1.7 Hz) is close to the frequency observed during the forced vibration tests performed in 1974 (1.77 Hz) as reported by Clinton et al. [2006].

3. Weak motion – Application to the Grenoble City Hall

The Grenoble City Hall building is a 13-story RC-structure built in 1967. The tower has a 44 m by 13 m plan section and rises 52 m above the ground. Two inner cores, consisting of RC shear walls, enclose the stairwells and lift shafts and are located at two opposite sides of the building. The structural strength system combines these shear walls with RC frames and longitudinal beams bearing the full RC floors. Since November 2004, the French Accelerometric Network (RAP) surveys the building [Péquegnat et al., 2008]. The accelerometric stations composed of 3C Episensors (Kinemetrics) and MiniTitan (Agecodagis) digitizers are localized at the top and the basement of the building. Michel [2007] showed that the building response under ambient vibrations computed using the Frequency Domain Decomposition method [Brincker et al., 2001] was dominated by the first bending modes in the longitudinal (1.16 Hz) and the transverse directions (1.22 Hz) and the first torsion mode (1.45 Hz).

The reassigned smoothed pseudo Wigner-Ville distribution has also been computed for the 3 earthquakes producing the largest vibrations at the OGH6 roof station (Fig. 2). Figure 2 shows the energy distribution between the bending and the torsion modes in the transverse direction. For the three examples given Fig. 2, only the bending mode (at 1.22 Hz) is activated during the time window corresponding to the strongest motion. Under ambient vibration (pre-event window) and for the Coda of the seismic signal, torsion mode (at 1.45 Hz) has energy, excepted for the Lago Di Garda ($M_L$=5.5 at 340 km, PGA=0.3 cm/s$^2$) earthquake. We observe also small and transient variations of the bending frequency (1.22 Hz) on short time windows of ambient vibrations (1%) and earthquakes (until 10%) (Fig. 2). The greatest variations occur for the Vallorcine ($M_L$=4.9 at 127 km) earthquake that produced a peak acceleration at the top of 11 cm/s$^2$.

In order to understand this transient variation, we developed a single degree-of-freedom model for this building, fixed using ambient vibration surveys done by Michel [2007]. We computed the synthetic time history at the roof by convolution of the recording time history at the basement with the single degree-of-freedom model using the Duhamel integral. Time frequency distribution applied to the experimental and synthetic time histories show identical frequency variations (Fig. 3). Because of the model used in this study is a 1D, linear and elastic model, these variations under weak motion must be only due to the input motion rather than stiffness variations in the building.

4. Discussion

It is well known that the structural response of civil engineering structures depends on three effects: the amplitude and frequency content of the loading, the soil-structure interaction and the dynamic properties of the structure.

In case of strong motions, the time-frequency distributions obtained for the San Fernando earthquake (Fig. 1) clearly show variations of the resonance frequencies from the pre- to co-seismic values and from the co- to post-seismic values. These variations are coherent with those estimated by different authors ([Clinton et al., 2006], [Dunand et al., 2006]) and due to the transient and permanent variations of the structural stiffness. In the case of the San Fernando earthquake, the co-seismic frequency value is reached quickly (5 to 10 s) whereas the post-seismic value is recovered slowly. The difference between the post- and pre-seismic frequencies is due to a permanent loss of building stiffness as firstly mentioned by Omori [1922] and followed by an abundant literature (e.g. [Stubbs and McLamore, 1973], [Udwadia and Trifunac, 1974], [Mucciarelli et al., 2004]) for other structures. The recovery between the co- and post-seismic frequencies show the transient variation of the stiffness that may be due to the re-closing process of cracks in the structure and opened during the greatest amplitudes of motion, or non-linear soil structure interaction. These two assumptions cannot be distinguished here and should be analysed in more details.

Figure 2 shows how the energy may jump from the bending to the torsion mode. Since the input motion is weak, that indicates that this variation is directly linked to the input motion instead of the structural parameters. The effect of the input motion is also observed on Fig. 3. In this case, the 1D linear model of the Grenoble City Hall used here reproduces the same transitory variation of the frequencies observed on the data. Even if some structural effects may exist, the input motion seems to dominate the frequency variation in the structure. Another aspect not detailed in this paper concerns the effect of the soil-structure interaction in the frequency variation. Recently, Clinton et al. [2006] and Todorovska and Al Rjoub [2006] observed the wander of building frequency due to the water contain in soil on long time series of seismic noise observation. The complexity of the input wavefield due to the wave propagation in the Earth and producing variations in amplitude and frequency may also explain the energy jump between modes and the transient variation of the apparent frequency observed in this study for short time windows.

Conclusion

Time-frequency distributions, and particularly the smoothed reassigned pseudo-Wigner-Ville method, allow understanding better the non-linear evolution of the resonance frequency of buildings during strong motions. We showed that a quick frequency drop occurs just after the peak acceleration until a minimum, followed by a slower increase in frequency until the post-seismic frequency. This frequency corresponds to those recorded under ambient or forced vibrations after the earthquake. In the case of the San Fernando earthquake recorded in the Millikan Library, a transient drop of 20 to 35% in N and E directions, respectively, were observed whereas the permanent drop was of 7 and 16.6%, respectively. In case of strong shaking like those produced by the San Fernando earthquake, the time-frequency distribution allowed therefore a relevant interpretation of the resonance frequency variations during earthquakes and then the assessment of the building integrity after damaging earthquake.

Under weak motion, the frequency variations observed on motion recorded at the City-Hall of Grenoble may be dominated by other effects such as variations of the frequency content of the incoming signal. In this study, the time-frequency analysis shows clearly the wander of the energy between the bending and the torsion modes. Since weak motions were used, these variations cannot be linked to non-linear effects in the structure or soil-structure interaction effects, as supported by the elastic and linear model applied to the City Hall of Grenoble.

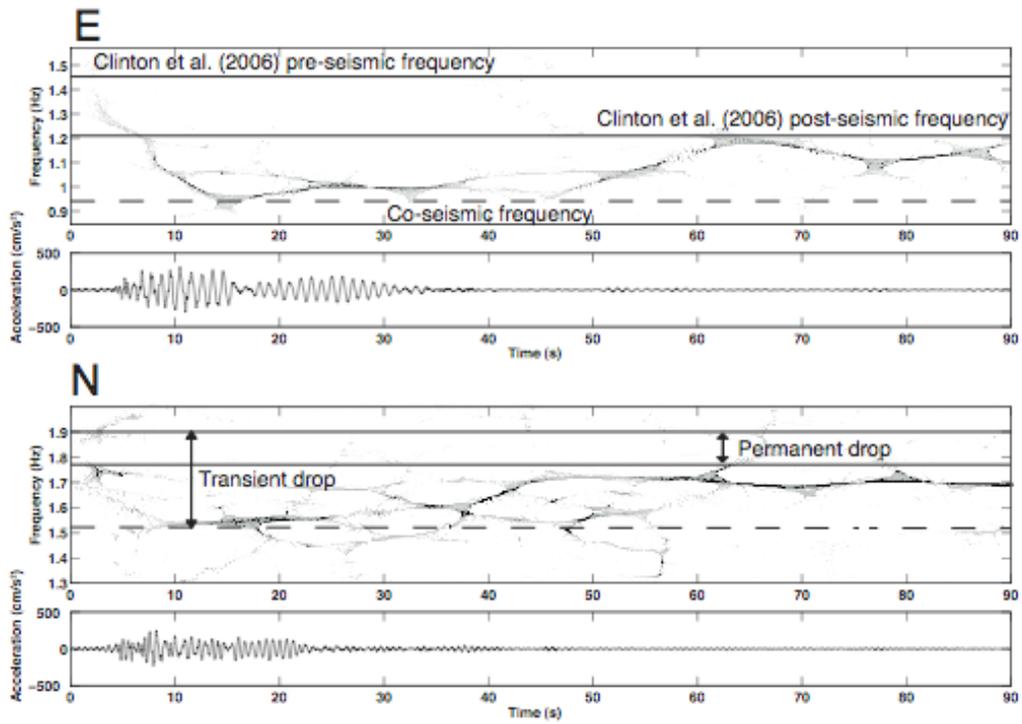

Figure 1: Time-Frequency distribution (smoothed reassigned pseudo-Wigner-Ville) of 1971/02/09 $M_L$=6.6 San Fernando Earthquake recordings at the roof of the Millikan Library on CalTech campus (California) in E (top) and N (bottom) directions.

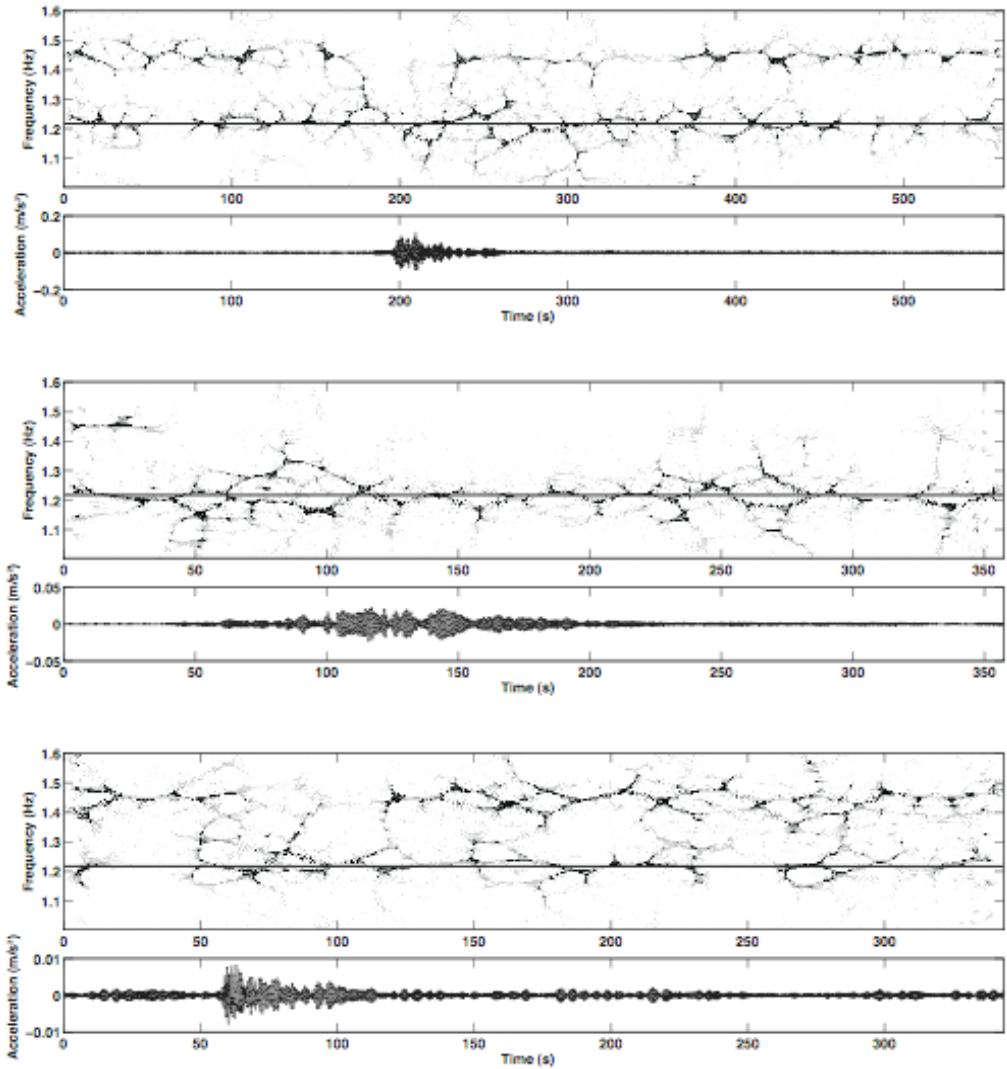

Figure 2: Time-Frequency distribution (smoothed reassigned pseudo-Wigner-Ville) of the recordings at OGH6 station in the transverse direction of Vallorcine Earthquake ($M_L$=4.9 Δ=127 km) (top), Lago di Garda Earthquake ($M_L$=5.5, Δ=340 km) (middle) and Laffrey Earthquake ($M_L$=3.1, Δ=15 km) (bottom). The horizontal black line corresponds to the first transverse resonance frequency obtained under ambient vibrations (1.22 Hz) (Michel, 2007). The first torsion mode frequency is 1.45 Hz.

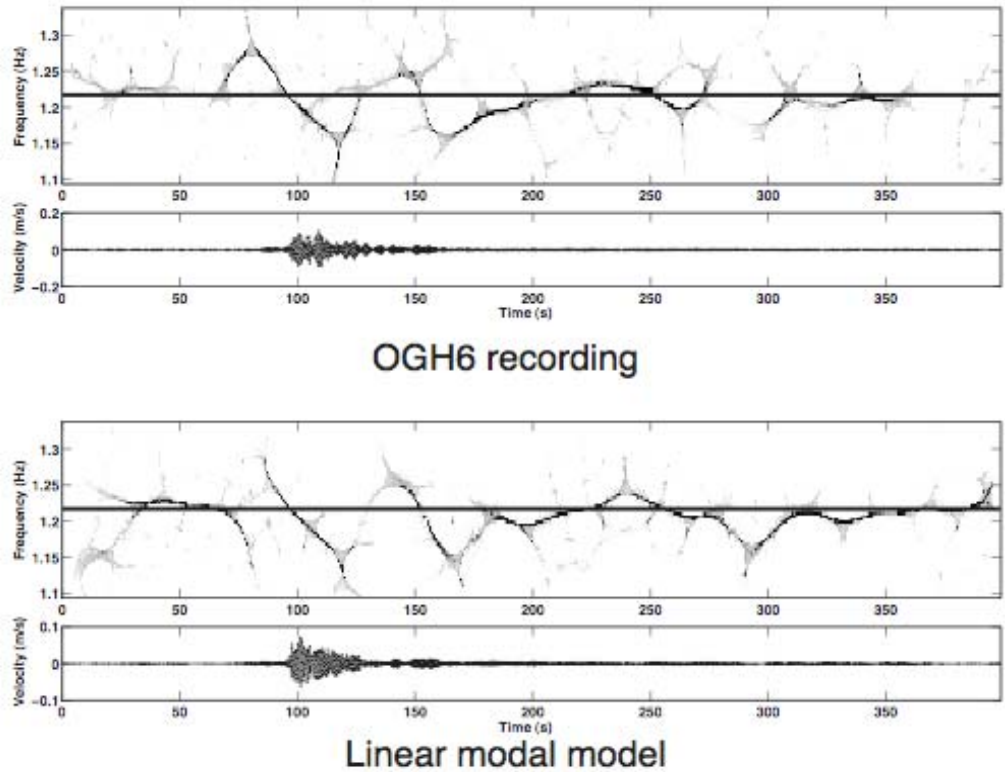

Figure 3: Comparison between time-frequency distribution (smoothed reassigned pseudo-Wigner-Ville) of the Vallorcine Earthquake ($M_L$=4.9, $\Delta$=127 km) OGH6 recording and the corresponding signal computed using the modal model of the City Hall based on modal parameters determined under ambient vibrations.